\documentclass[aps,prb,groupedaddress,twocolumn,showpacs]{revtex4}
\usepackage{graphicx}
\usepackage{multirow}

\begin{document}
\title{Critical behavior of dynamic vortex Mott transition in superconducting arrays at fractional vortex densities.}

\author{Enzo Granato}

\address{Laborat\'orio Associado de Sensores e Materiais,
Instituto Nacional de Pesquisas Espaciais, 12227-010 S\~ao Jos\'e dos
Campos, SP, Brazil}

\begin{abstract}
We study the differential resistivity transition of two-dimensional superconducting arrays induced by an external driving current, in 
the presence of thermal fluctuations and a magnetic field corresponding to $f$ flux quantum per plaquette.
Recent experiments have identified this transition as a dynamic vortex Mott insulator transition at vortex densities near rational values of  $f$. 
The critical behavior is determined  from a scaling analysis of the current-voltage relation near the transition, obtained
by Monte Carlo simulations of a Josephson-junction array model in the vortex representation. For a square-lattice array, the critical exponents obtained near $f=1/2$ are consistent with the experimental observations. The  same scaling behavior is observed near  $f=1/3$. For a honeycomb array, although similar results are obtained for $f=1/3$,  the transition is absent for $f=1/2$, consistent with an incommensurate vortex phase.

\end{abstract}
\pacs{74.81.Fa, 74.25.Uv}

\maketitle

\section{Introduction}

The concept of a Mott insulator phase, which has been of fundamental importance for the understanding of  transport properties of many materials with strong Coulomb interactions,  can also be extended to other quantum systems \cite{fisherweich1989,fishergrinst1990,greiner2002} and even to systems without quantum fluctuations \cite{poccia15,rade17,nelsonvinokur93}. An important case is the vortex lattice in type II superconductors in presence of columnar pinning defects, when the vortex density is close to the density of pinning sites \cite{nelsonvinokur93}.  Below a critical temperature,  the vortex mobility vanishes because the vortex occupied pinning sites prevent motion of  individual vortices, leading to a zero resistance state. The analogy to a Mott insulator phase follows from the  mapping of the three-dimensional vortex lattice system to  interacting bosons in two dimensions at zero temperature, where there is an insulating phase of localized bosons with a finite energy gap. Transitions out of the Mott insulator phase have been widely studied.  Besides thermodynamic transitions as a function of temperature \cite{limelette2003,rozenberg1999},  dynamic Mott transitions induced by an external driving current have attracted particular attention \cite{poccia15,lankhorst2018,rade17,li2015,stoliar2013}, recently. Since the dynamic Mott transition appears to have similar critical behavior as the thermal one \cite{poccia15,limelette2003}, with the temperature corresponding to the current, studies of the  dynamical version might help to understand the critical behavior of the equilibrium transition.   

Recently, a remarkable dynamic vortex Mott transition has been revealed through experiments on two-dimensional Josephson-junction arrays (JJA) in the form of  superconducting grains coupled  by the  proximity effect  on a square lattice, in the presence of a perpendicular magnetic field  \cite{poccia15,lankhorst2018}. It corresponds to a differential resistivity transition at low  temperatures and  finite driving currents, instead of the usual equilibrium resistivity transition at higher temperatures and zero current due to thermal fluctuations  \cite{teiteljaya83,shih1985,newrock2000,eg08,eg13} or the dynamical depinning transitions at higher currents and low temperatures  \cite{marconi2001}.
For a JJA in an external magnetic field, the average vortex density is determined by the frustration parameter $f$, corresponding to the number of flux quantum per plaquette. The equilibrium phase transitions of a JJA are strongly dependent \cite{teiteljaya83} on the value of $f$ and the geometry of the lattice \cite{shih1985}. While for a square lattice, rational values of $f$ leads to vortex lattices commensurate with the underlying pinning potential and finite temperature resistive transitions, in a honeycomb lattice with $f=1/2$, the existence or nature of the equilibrium transition is not fully understood, due to an additional geometric frustration \cite{eg13}.  In the recent experiments on a square lattice \cite{poccia15}, dynamic vortex Mott insulator to metal transitions were clearly identified near rational vortex densities, such as  $f=1, 2 $ and $1/2$. The differential resistivity as a function of $f$ for increasing currents displays the reversal of a minimum into a maximum near these  values of $f$.
This is the analog of  the dip-to-peak reversal of  the electronic density of states near the Fermi level of the (quantum) Mott insulator to metal transition \cite{lankhorst2018,rozenberg1999}. Similar behavior was observed earlier in other superconducting arrays but it was not regarded as a manifestation of a dynamic transition \cite{benz90,jiang2004}. The scaling behavior of the differential resistivity as a function of the deviation of current $\delta I$ and  frustration $\delta f$ from their critical values $I_c$ and $f_c$, was shown to be described by a single critical exponent $\epsilon$. For  $f_c=1$ and $ 2$, the obtained value, $\epsilon = 2/3$, is consistent with a mean field description of phase slip dynamics \cite{poccia15} and, more recently, with results obtained by mapping this dynamical transition into a non-Hermitian quantum problem \cite{tripathi2016,lankhorst2018}.  On the other hand, near $f_c=1/2$, a distinct critical exponent $\epsilon =0.5$ was found, indicating  that the dynamic vortex Mott transitions at fractional vortex densities may belong to different universality classes. Very recent Monte Carlo (MC) simulations for a model of particles with long-range Coulomb interactions \cite{rade17} near a particle density $1/2$
found a very different critical exponent, $\epsilon=1.5$,  which might be due to the form of the interaction potential.  Therefore, a more realistic JJA array model, where  vortices  interact logarithmically, is required to clarify this interesting question and also investigate the effects of lattice geometry.

In this work we study the differential resistivity transition (dynamic Mott transition) in superconducting arrays by MC simulations of a JJA model in the vortex representation with logarithmic interactions.  The critical behavior is determined  from a scaling analysis of the current-voltage characteristics and differential resistivity near the transition.  From scaling arguments, we obtain  $\epsilon=1/2 \nu$, where $\nu$  is the correlation length exponent.  For a square lattice, we find $\epsilon \sim 0.5$ near frustration $f=1/2$, consistent with the experimental observations. The  same scaling behavior is observed near  $f=1/3$. For a honeycomb array, although similar results are obtained for $f=1/3$,  the transition is absent for $f=1/2$, consistent with an incommensurate vortex phase.

\section{Model and simulation}

We consider two-dimensional superconducting arrays in a transverse magnetic field, described by the JJA Hamiltonian \cite{teiteljaya83,newrock2000}
\begin{equation}
\mathcal{H} = -E_o \sum_{<k l  >} \cos (\theta_k -\theta_{l} - A_{k l }  ),
\end{equation}
where $\theta_k$ is the phase of the local superconducting order parameter at the sites $k$ of the lattice, $E_o=(\hbar/2 e) I_o$ and
$I_o$ is the single-junction critical current. 
The line integral of the vector potential $A_{k l }$
due to the external field $\vec B = \nabla \times \vec A$ is
constrained to $\sum_{k l}A_{k l} = 2 \pi f$ around each elementary
plaquette, where $f$ is the number of flux quantum $\phi_o=hc/2e$
per plaquette. This model is periodic in $f$ with period $f = 1$ with reflection symmetry about $f=1/2$.
To study the vortex dynamics it is convenient to rewrite
the above phase Hamiltonian in the vortex representation 
\begin{equation}
H = 2 \pi^2 E_o  \sum_{i,j} (n_i- f) G^\prime_{i,j} (n_j -f)   ,
\label{cghamilt}
\end{equation}
which  can  be  obtained  following  a  standard  procedure \cite{jose1977}
in which the phase model is replaced by a periodic Gaussian model, leading to explicit vortex variables 
represented by integer charges $n_i$ at the sites $  r_i=(x_i,y_i)$ of the dual lattice and 
constrained by the neutrality condition, $\sum_i (n_i -f )=0$.
The vortex interaction is given by $G^\prime_{ij} = G( r_i - r_j) - G(0)$, where  $G( r)$ is 
the lattice Green's function corresponding to the dual lattice of the JJA geometry \cite{franz1995vortex,leeteitel94,hyman95}, which
is square and triangular for square and honeycomb arrays, respectively.   $G^\prime (\bf  r) $  
diverges logarithmically as  $ -\log(r)/(2 \pi) $ for large separations.  
 
For a square lattice,
\begin{equation}
G({\bf  r}) =\frac{1}{L^2} \sum_k \frac{e^{i {\bf k}\cdot  {\bf r} }}{4-2 \cos({\bf k}\cdot {\bf a_1}) - 2 \cos({\bf k}\cdot {\bf a_2}) },
\end{equation}
where $L$ is the system size, $\bf  k$ are the reciprocal lattice vectors and $\bf a_1$, $\bf a_2$ are two perpendicular nearest-neighbor lattice vectors.   
For a JJA on a honeycomb lattice, the dual lattice is triangular and the corresponding lattice Green's function is given by \cite{franz1995vortex}
\begin{equation} 
G({\bf r}) =\frac{1}{2 L^2} \sum_k \frac{e^{i {\bf  k}\cdot {\bf r} }}{3- \cos({\bf k}\cdot {\bf a_1}) -  \cos({\bf k}\cdot {\bf a_2}) -\cos({\bf k}\cdot {\bf a_3})},
\end{equation}
where ${\bf a_1}, {\bf a_2} $, and ${\bf a_3}$ are three nearest-neighbor lattice vectors separated by $120^\circ $ from each other.

We  study  the  nonequilibrium  response  of  the JJA under an applied driving current
by  driven MC  simulations  of  the vortex  model under an applied force \cite{leeteitel94,hyman95,gK98,eg98}. 
The vortex dynamics is assumed to be ovedamped. 
The force  represents the effect of the driving current density $J$ on the vortices, acting as  a Lorentz force transverse to
the velocity, leading to  an   additional   contribution   to   the   energy   in   Eq.  (\ref{cghamilt}),  $- (h/2 e) J  \sum_i n_i x_i $, 
when $J$ is  in the $\hat y$ direction.  The  MC  time  is
identified as the real time $t$ with the unit of time $dt=1$, corresponding to a complete MC pass through
the lattice. A MC step consists of adding a dipole of vortex charges to a nearest-neighbor charge pair $(n_i, n_j)$, 
using the Metropolis algorithm. Choosing a nearest-neighbor  pair $(i,j)$ at random, the step consists of changing  $n_i \rightarrow n_i -1 $ and $n_j \rightarrow n_j +1 $,
corresponding to the motion of a unit charge by a unit length from $ r _i$ to $  r_j$.  The move is accepted with
probability $ min[1,\exp(-\Delta H/kT) ] $, where  $\Delta H$ is the change in the energy. Periodic  boundary   conditions  
are used in systems of linear size $L$. The driving current $J$ biases the
added dipole, leading to a  net flow of vortices in the direction transverse to the current, if the vortices are mobile.  This vortex flow generates an electric field $E$ along the current which can be calculated (in arbitrary units) as $E(t)= \frac{1}{L} \sum_i \Delta Q_i(t) $,
after each MC pass through the lattice, where  $\Delta Q_i=({\bf r_i} - {\bf r_j})\cdot {\bf \hat x} $ for an accepted vortex dipole excitation at the nearest-neighbor sites $(i,j)$ 
and  $\Delta Q_i=0$  otherwise.  Due to the neutrality condition, $f$ is varied 
in multiples of $1/L^2$. Temperature $T$ is measured in units of $E_o$ and $J$ in units of $I_o/2 \pi$. 
We use typically $4 \times 10^5$ MC passes to compute time averages and the same number of passes to reach steady states. 

\section{Results and discussion}

\subsection{Differential resistivity transition}

\begin{figure}
\centering
\includegraphics[width=\columnwidth]{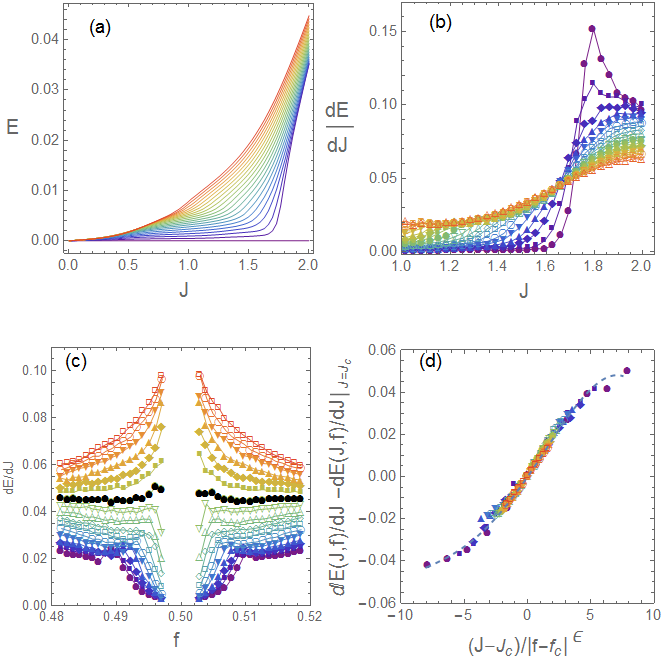}
\caption{ (a) Current-voltage (I-V) relation ($J=I/L$, $E=V/L$  ) for increasing frustration $f$ near $f=1/2$ for a JJA on a square lattice. Temperature $T=0.2$ and system size $L=32$. From the bottom up,  $f$ increases from $0.5$ to $0.5195$ in $19$ equal steps. (b) Differential resistivity $\frac{d E}{d J}$ for $f>1/2$ obtained numerically from (a) near the current induced dynamical transition. (c) $\frac{d E }{d J}$  plotted as a function of $f$  for different driving currents. From the bottom up, $J$ increases from $1.322 $ to $1.865 $ in $16$ equal steps. Black dots indicate  the separatrix $d E/d J|_{J=J_c}$.
(d) Scaling plot of $\frac{d E}{dJ}$  near the dynamical transition for $f > 1/2$, with  $J_c=1.66$, $ f_c=0.502$, and $\epsilon=0.55$. }
\label{ivf12}
\end{figure}

\begin{figure}
\centering
\includegraphics[width=\columnwidth]{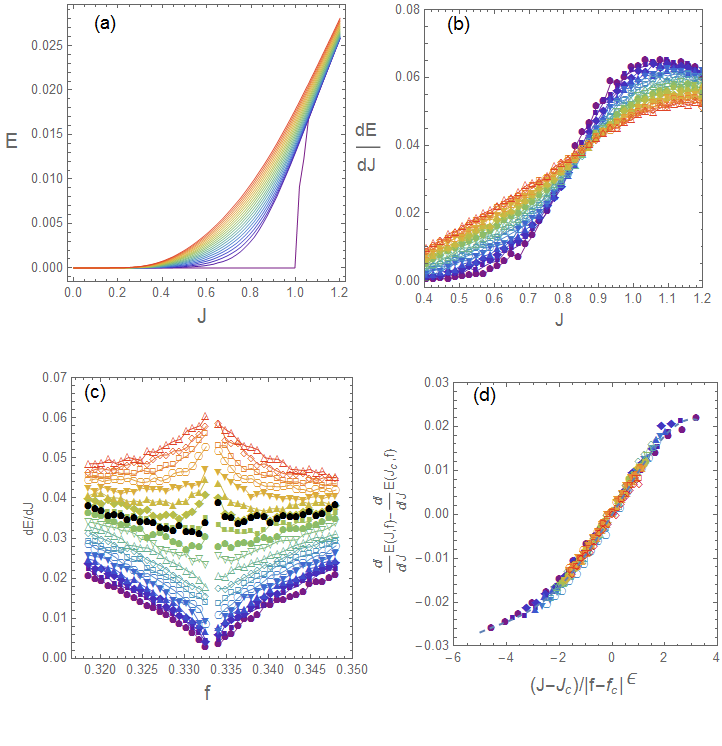}
\caption{Same as Fig. \ref{ivf12} but for $f=1/3$  and $L=36$. (a) From the bottom up,  $f$ increases from $1/3$ to $0.3480$ in 20 equal steps.
(b) $\frac{d E}{d J}$ for $f > 1/3$  near the transition. (c) $\frac{d E }{d J}$  as a function of $f$  for different $J$. From the bottom up, $J$ increases from $0.5898 $ to $0.9559 $ in $19$ equal steps. 
(c) Scaling plot of $\frac{d E}{dJ}$  for $f > 1/3$, with  $J_c=0.83$, $ f_c=1/3$ and $\epsilon=0.5$.}
\label{ivf13}
\end{figure}

We first consider the dynamical transition near $f=1/2$ for a JJA on a square lattice. The effect of  increasing the frustration $f$ from $f=1/2$ on the current-voltage (I-V) relation is shown in Fig. \ref{ivf12}a, in terms of the current density $J=I/L$ and electric field $E=V/L$. The temperature is much below the critical temperature, $T_c \approx 0.8$, of the equilibrium resistive transition ($J=0$) of the corresponding commensurate vortex lattice \cite{leeteitel94,gK98} at $f=1/2$ and so the linear resistivity  $\rho_L= d E/ d J |_{J=0}$ vanishes. The range of currents is also much below the corresponding zero-temperature depinning current, $J_p\approx 4.9$. While $E$ remains essentially zero for increasing $J$ when  $f=1/2$, a small increment in $f$ leads to a sharp increase above a critical value $J_c\approx 1.6-1.7$ [Fig. \ref{ivf12}a]. Further increase of $f$ tends to  smooth out the slope of the  $E \times J$ curve near $J_c$. This change in the slope can be seen much clearer in the behavior of the differential resistivity, $dE/dJ$, as shown in Fig. \ref{ivf12}b, which also reveals that the
curves for different $f$ above $f_c \approx 1/2$ cross approximately at the same point  $J_c$. The crossing point strongly suggests the presence of an  underlying continuous transition, where $dE/dJ$ behaves as  a scaling invariant quantity and $f - f_c$ acts as a relevant perturbation. In such a case, one expects the scaling behavior $dE/dJ  \approx F(\delta J/|\delta f|^\epsilon)$, where $F(x)$ (with $ F(0)=c$, a constant) is a scaling function,  $\delta J=J-J_c$, $\delta f = f -f_c$ and $1/\epsilon$ is the  crossover critical exponent. Alternatively, $dE/dJ$ plotted as a function of $f$ for different currents shown in Fig. \ref{ivf12}c, where data for $f < 1/2$ is also included,  clearly shows the reversal of a minimum into a maximum near $f=1/2$ for increasing current density  at  $J_c$. This behavior was first observed in the experiments and identified as a signature of the dynamic vortex Mott insulator-metal transition \cite{poccia15}. Since $J_c$ is well below the depinning critical current, this transition is unrelated to crossover effects near the current-induced vortex lattice depinning.  To verify the scaling behavior, we plot the data near the transition, in Fig. \ref{ivf12}d, according to a similar scaling form, which was proposed in the experiments \cite{poccia15}, 
\begin{equation}
d E /d J - d E/d J|_{J=J_c} = H(\delta J/|\delta f |^\epsilon),
\label{exp}
\end{equation}
where $H(0)=0$. Data for different $J$ and $f$  collapse into the same smooth curve when $J_c$, $f_c$ and $\epsilon $ have the appropriate values. 
The scattering of the data is mainly due to the numerical calculation of the derivative $dE/dJ$. 
The data collapse was obtained by varying $J_c$ and $f_c$ slightly from the above estimates and using different values of $\epsilon$ to estimate an errorbar. 
The value obtained for the critical exponent,  $\epsilon = 0.50(6)$, is consistent with the one obtained from the experiments, strongly supporting the universality of this dynamical transition. Our numerical results, which were  obtained for a model of logarithmically interacting vortices, also supports the recent conjecture \cite{rade17} that this exponent depends on the form of the interaction potential,  since $\epsilon =1.5$  was found  for a model of particles with long-range Coulomb interactions.
Another low-order commensurate frustration, $f=1/3$, displays the same scaling behavior with $\epsilon = 0.50(6)$ (Fig. \ref{ivf13}). 

\begin{figure}
\centering
\includegraphics[width=\columnwidth]{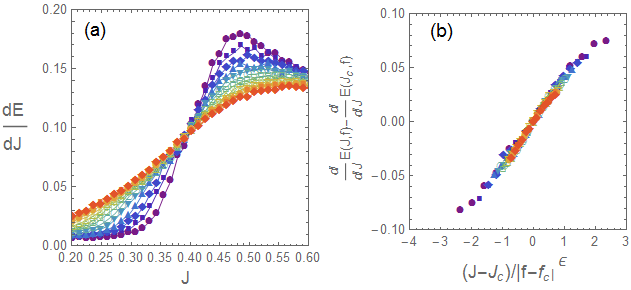}
\caption {(a)  Differential resistivity $\frac{d E}{d J}$ near $f=1/3$ for a JJA on a honeycomb lattice near the current induced dynamical transition. Temperature $T=0.2$ and system size $L=36$. $f$ increases from $0.3387$
to $0.3480$ in $13$ equal steps.
(b)  Scaling plot of $\frac{d E}{dJ}$  near the dynamical transition with  $J_c=0.403$, $ f_c=0.3378$, $\epsilon=0.5$. }
\label{ivf13H}
\end{figure}

\begin{figure}
\centering
\includegraphics[width=\columnwidth]{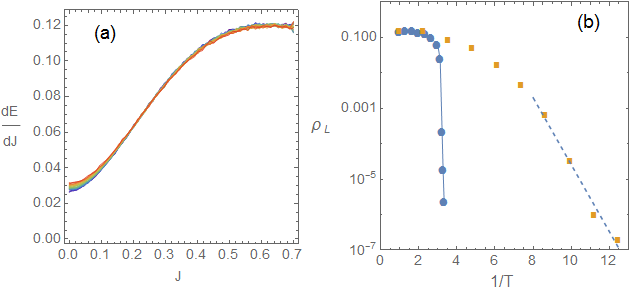}
\caption{(a) Differential resistivity $\frac{d E}{d J}$ for increasing frustration $f$ near $f=1/2$ for a JJA on a honeycomb lattice.  Temperature $T=0.2$ and system size $L=36$. $f$ increases from $0.5$ to $0.5131$ in $18$ equal steps. (b) Linear resistivity $\rho_L$ as function of temperature $T$ for $f=1/2$ (squares) and $f=1/3$ (circles).  Dotted line corresponds to an Arrhenius thermal activated behavior.  }
\label{ivf12H}
\end{figure}

\begin{figure}
\centering
\includegraphics[width=\columnwidth]{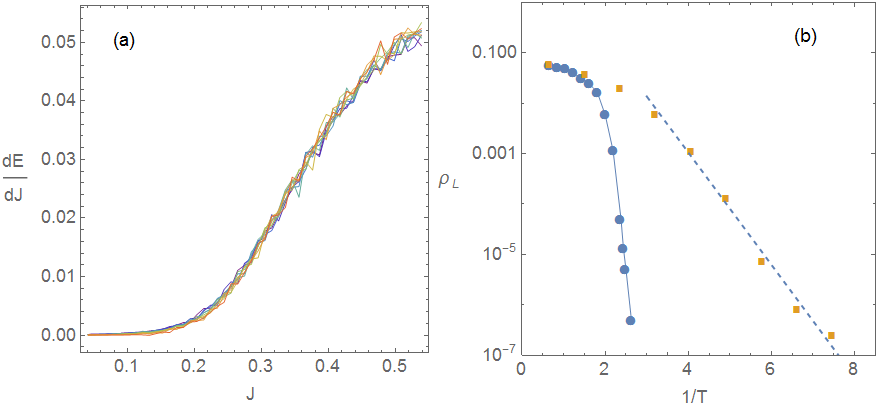}
\caption{(a) Differential resistivity $\frac{d E}{d J}$ for increasing magnetic frustration $f$ near an irrational frustration $f=(3-\sqrt{5})/2$ (Golden ratio) for a JJA on a square lattice. Temperature $T=0.2$ and system size $L=34$.  $f$ increases from $0.3824$ to $0.3901$ in $9$ equal steps. (b) Linear resistivity $\rho_L$ as function of temperature $T$ for $f$ irrational (squares) and $f=1/3$ (circles).  Dotted line corresponds to an Arrhenius thermal activated behavior.  }
\label{ivirrat}
\end{figure}

For a JJA on a honeycomb lattice, similar behavior is also found for  $f=1/3$ (Fig.  \ref{ivf13H}). 
However, for $f=1/2$, there is no crossing in the current dependence of $ d E/ d J $ for increasing frustration (Fig. \ref{ivf12H}a), indicating the absence of a dynamic vortex Mott transition. Moreover, since  $ d E/ d J |_{J\rightarrow 0}$ is nonzero, the linear resistivity $\rho_L$  is finite. To confirm this behavior, we also obtained $\rho_L$ from equilibrium voltage fluctuations, without imposing a current bias, using the fluctuation-dissipation relation 
\begin{equation}
\rho_L=\frac{1}{2 kT} \int dt < V(0) V(t)>.
\end{equation}
As shown in Fig. \ref{ivf12H}b, the linear resistivity remains finite at lower temperatures, displaying an Arrhenius behavior while for $f=1/3$ it vanishes below a critical temperature. This is consistent with the absence of an equilibrium resistive  transition at finite temperatures for $f=1/2$ on a honeycomb lattice, in agreement with recent numerical simulations for the same model in the phase representation \cite{eg13}. 
For the JJA on square lattice at irrational frustration,  $f=(3-\sqrt{5})/2$ (Golden ratio),  there is also no crossing in the current dependence of $ d E/ d J $ for increasing frustration and so a dynamical transition is also absent (Fig. \ref{ivirrat}). The linear resistivity  $\rho_L$ remains finite at lower temperatures, displaying an Arrhenius behavior while for $f=1/3$ it vanishes below a critical temperature. The absence of an equilibrium transition 
for irrational $f$ is also in agreement with  simulations in the phase
representation \cite{eg08,eg96}.  Therefore, an underlying equilibrium resistive transition at nonzero temperatures is required for the observation of  the dynamical transition at nonzero driving currents at lower temperatures.

\subsection{Current-voltage scaling}

\begin{figure}
\centering
\includegraphics[width=\columnwidth]{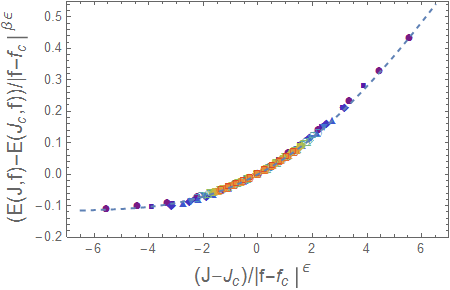}
\caption { Scaling plot of $E(J,f)$ near the dynamical transition for $f=1/2$ on a square lattice, with  $J_c=1.66$, $ f_c=0.502$, $\beta =1 $ and  $\epsilon=0.5$. Temperature $T=0.2$ and system size $L=32$}
\label{ivscaling}
\end{figure}

We now describe the expected behavior of the differential resistivity from general arguments of the dynamic scaling theory for the current-voltage characteristics  \cite{hyman95}.  Assuming a continuous dynamical transition at $J_c$, measurable quantities should scale  with the diverging correlation length  $\xi \sim |\delta J|^{-\nu} $ and relaxation time $\tau \sim \xi^{z}$, where $\nu$ and $z$ are the correlation length and dynamic critical exponents, respectively.  Since the electric field $E$ generated by moving vortices with density $f$ and velocity $v$  is proportional to  $ f v$,  the singular contribution to $E$ should scale as $E \sim \xi^{1-z}$. Crossover effects due to a change  $\delta f$ should occur when  $ |\delta f |\xi^2 \approx 1$, corresponding to an additional vortice in a correlated area, revealing that $\delta f$ is a strongly relevant perturbation and should therefore appear in the scaling function in the combination $\delta J/|\delta f|^{\epsilon}$, with $\epsilon=1/2 \nu$.   As a function of $\delta J$ and $\delta f$,  one then expect the scaling behavior
\begin{equation}
E(J,f) = F_o(J,f) + |\delta J|^\beta F_1(\delta J /|\delta f|^\epsilon),
\label{IVscal}
\end{equation}
where $\beta = (z-1 )\nu $, $F_o$ is a regular contribution, analytic in $\delta J$ and $\delta f$, and $F_1(x)$ is a scaling function with $F_1(0)=c$, a constant. The scaling form for the differential resistivity  $dE/dJ$ can then be written as
\begin{equation}
\frac{dE(J,f)}{d J}- \frac{dE(J,f)}{d J}|_{J=J_c}= |\delta f|^{(\beta -1)\epsilon} H(\delta J /|\delta f|^\epsilon),
\label{diffscal}
\end{equation}
with $H(0)=0$. We have neglected the $\delta J$ dependence of  $ \frac{dF_o(J,f)}{d J}$. This scaling form reduces 
to the one used in the experiments (Eq. \ref{exp}) when $\beta=1$, which can be obtained with $z=2$ and  $\nu=1$, leading to a crossover exponent  $\epsilon=1/2 \nu =0.5$, in agreement with the experimental results \cite{poccia15} and the data collapse for $dE/dJ$ in Fig. \ref{ivf12}d. In fact, a good data collapse is obtained for the bare data $E(J,f)$ with these critical exponents, according to the scaling form of Eq. \ref{IVscal} (Fig. \ref{ivscaling}). 

\begin{figure}
\centering
\includegraphics[width=\columnwidth]{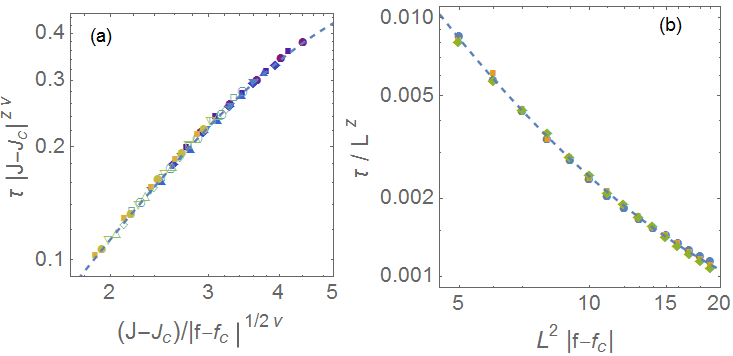}
\caption{ Scaling plots of the relaxation time  $\tau$  near the dynamical transition for $f = 1/2$ on a square lattice
(a) Near the transition for increasing $f$ and $J >J_c$ with  $J_c=1.54$, $f_c=0.502$, $z=2$ and $\nu=1$.   $f$ increases from $0.5035$ to $0.5083$ in $11$ equal steps.
(b) At the transition for increasing $f$ and  different system sizes, $L=42$, $48$, and $ 54 $  with $J_c=1.61$, $f_c=0.502$ and $z=2.1$. }
\label{tau}
\end{figure}

To check the conjectured values of  $z$ and $\nu$ from independent data, we performed a  scaling analysis of the relaxation time $\tau(J,f)$, obtained from the voltage time correlation function  
\begin{equation}
C(t)= \frac{<V(t) V(0)> - <V(t)>^2}{(<V(t)^2>-<V(t)>^2} . 
\end{equation}
Near the transition, $\tau$ can be estimated from the expected time dependence of $C(t)$ at long times, 
$C(t)\propto e^{-t/\tau}$.
Using $\epsilon=1/2 \nu$,  $\tau$  should then satisfy the  scaling form 
\begin{equation}
\tau  |\delta J|^{z \nu} = F_2(\delta J /|\delta f|^{1/2 \nu}) ,
\end{equation}
in absence of finite-size effects. At the transition, the correlation length is cutoff by the system size $L$ and $\tau$ should satisfy the finite-size scaling form
\begin{equation}
\tau/ L^z  = F_3(L^2|\delta f|) .
\end{equation}
Indeed, as shown in Figs.  \ref{tau}a and \ref{tau}b, a reasonable data collapse  according to the above  scaling forms  are obtained with $z\approx 2$ and $\nu\approx 1$.

\section{Conclusions}

In summary, we have studied a current-induced dynamical transition in two-dimensional JJA, with a sharp change in the differential resistance behavior, which is a manifestation of the dynamic vortex Mott insulator transition identified in experiments \cite{poccia15,lankhorst2018}. 
From  MC simulations and scaling analysis, we find  the critical exponent $\epsilon = 0.50(6)$ near $f=1/2$, consistent with the experimental observations for a square array and the conjecture \cite{rade17} that this exponent depends on the form of the interaction potential.  For a honeycomb array, however,  the transition is absent for $f=1/2$.  As a consequence,  minimum to maximum reversal in the magneto differential resistance for increasing current should not occur for $f=1/2$  in such systems, as for example, in superconducting thin films with a triangular lattice of nanoholes \cite{valles07,roy2017}, which can be modeled by a honeycomb JJA \cite{eg13}. It should be noted that these results were obtained  assuming overdamped vortex dynamics. Nevertheless, it should be  interesting to study the effects of underdamped  dynamics, as this approximation may not be appropriate  for JJA in general.

\acknowledgements


The author thanks V. M. Vinokur,  T. I. Baturina and J.M. Valles for helpul discussions. Work supported by CNPq (Conselho Nacional de Desenvolvimento Cient\'ifico e Tecnol\'ogico)  and  CENAPAD-SP.


\begin{thebibliography}{29}
\expandafter\ifx\csname natexlab\endcsname\relax\def\natexlab#1{#1}\fi
\expandafter\ifx\csname bibnamefont\endcsname\relax
  \def\bibnamefont#1{#1}\fi
\expandafter\ifx\csname bibfnamefont\endcsname\relax
  \def\bibfnamefont#1{#1}\fi
\expandafter\ifx\csname citenamefont\endcsname\relax
  \def\citenamefont#1{#1}\fi
\expandafter\ifx\csname url\endcsname\relax
  \def\url#1{\texttt{#1}}\fi
\expandafter\ifx\csname urlprefix\endcsname\relax\def\urlprefix{URL }\fi
\providecommand{\bibinfo}[2]{#2}
\providecommand{\eprint}[2][]{\url{#2}}

\bibitem[{\citenamefont{Fisher et~al.}(1989)\citenamefont{Fisher, Weichman,
  Grinstein, and Fisher}}]{fisherweich1989}
\bibinfo{author}{\bibfnamefont{M.~P.} \bibnamefont{Fisher}},
  \bibinfo{author}{\bibfnamefont{P.~B.} \bibnamefont{Weichman}},
  \bibinfo{author}{\bibfnamefont{G.}~\bibnamefont{Grinstein}},
  \bibnamefont{and} \bibinfo{author}{\bibfnamefont{D.~S.}
  \bibnamefont{Fisher}}, \bibinfo{journal}{Physical Review B}
  \textbf{\bibinfo{volume}{40}}, \bibinfo{pages}{546} (\bibinfo{year}{1989}).

\bibitem[{\citenamefont{Fisher et~al.}(1990)\citenamefont{Fisher, Grinstein,
  and Girvin}}]{fishergrinst1990}
\bibinfo{author}{\bibfnamefont{M.~P.} \bibnamefont{Fisher}},
  \bibinfo{author}{\bibfnamefont{G.}~\bibnamefont{Grinstein}},
  \bibnamefont{and} \bibinfo{author}{\bibfnamefont{S.}~\bibnamefont{Girvin}},
  \bibinfo{journal}{Physical Review Letters} \textbf{\bibinfo{volume}{64}},
  \bibinfo{pages}{587} (\bibinfo{year}{1990}).

\bibitem[{\citenamefont{Greiner et~al.}(2002)\citenamefont{Greiner, Mandel,
  Esslinger, H{\"a}nsch, and Bloch}}]{greiner2002}
\bibinfo{author}{\bibfnamefont{M.}~\bibnamefont{Greiner}},
  \bibinfo{author}{\bibfnamefont{O.}~\bibnamefont{Mandel}},
  \bibinfo{author}{\bibfnamefont{T.}~\bibnamefont{Esslinger}},
  \bibinfo{author}{\bibfnamefont{T.~W.} \bibnamefont{H{\"a}nsch}},
  \bibnamefont{and} \bibinfo{author}{\bibfnamefont{I.}~\bibnamefont{Bloch}},
  \bibinfo{journal}{Nature} \textbf{\bibinfo{volume}{415}}, \bibinfo{pages}{39}
  (\bibinfo{year}{2002}).

\bibitem[{\citenamefont{Poccia et~al.}(2015)\citenamefont{Poccia, Baturina,
  Coneri, Molenaar, Wang, Bianconi, Brinkman, Hilgenkamp, Golubov, and
  Vinokur}}]{poccia15}
\bibinfo{author}{\bibfnamefont{N.}~\bibnamefont{Poccia}},
  \bibinfo{author}{\bibfnamefont{T.~I.} \bibnamefont{Baturina}},
  \bibinfo{author}{\bibfnamefont{F.}~\bibnamefont{Coneri}},
  \bibinfo{author}{\bibfnamefont{C.~G.} \bibnamefont{Molenaar}},
  \bibinfo{author}{\bibfnamefont{X.~R.} \bibnamefont{Wang}},
  \bibinfo{author}{\bibfnamefont{G.}~\bibnamefont{Bianconi}},
  \bibinfo{author}{\bibfnamefont{A.}~\bibnamefont{Brinkman}},
  \bibinfo{author}{\bibfnamefont{H.}~\bibnamefont{Hilgenkamp}},
  \bibinfo{author}{\bibfnamefont{A.~A.} \bibnamefont{Golubov}},
  \bibnamefont{and} \bibinfo{author}{\bibfnamefont{V.~M.}
  \bibnamefont{Vinokur}}, \bibinfo{journal}{Science}
  \textbf{\bibinfo{volume}{349}}, \bibinfo{pages}{1202} (\bibinfo{year}{2015}).

\bibitem[{\citenamefont{Rademaker et~al.}(2017)\citenamefont{Rademaker,
  Vinokur, and Galda}}]{rade17}
\bibinfo{author}{\bibfnamefont{L.}~\bibnamefont{Rademaker}},
  \bibinfo{author}{\bibfnamefont{V.~M.} \bibnamefont{Vinokur}},
  \bibnamefont{and} \bibinfo{author}{\bibfnamefont{A.}~\bibnamefont{Galda}},
  \bibinfo{journal}{Scientific Reports (Nature Publisher Group)}
  \textbf{\bibinfo{volume}{7}}, \bibinfo{pages}{44044} (\bibinfo{year}{2017}).

\bibitem[{\citenamefont{Nelson and Vinokur}(1993)}]{nelsonvinokur93}
\bibinfo{author}{\bibfnamefont{D.~R.} \bibnamefont{Nelson}} \bibnamefont{and}
  \bibinfo{author}{\bibfnamefont{V.}~\bibnamefont{Vinokur}},
  \bibinfo{journal}{Physical Review B} \textbf{\bibinfo{volume}{48}},
  \bibinfo{pages}{13060} (\bibinfo{year}{1993}).

\bibitem[{\citenamefont{Limelette et~al.}(2003)\citenamefont{Limelette,
  Georges, J{\'e}rome, Wzietek, Metcalf, and Honig}}]{limelette2003}
\bibinfo{author}{\bibfnamefont{P.}~\bibnamefont{Limelette}},
  \bibinfo{author}{\bibfnamefont{A.}~\bibnamefont{Georges}},
  \bibinfo{author}{\bibfnamefont{D.}~\bibnamefont{J{\'e}rome}},
  \bibinfo{author}{\bibfnamefont{P.}~\bibnamefont{Wzietek}},
  \bibinfo{author}{\bibfnamefont{P.}~\bibnamefont{Metcalf}}, \bibnamefont{and}
  \bibinfo{author}{\bibfnamefont{J.}~\bibnamefont{Honig}},
  \bibinfo{journal}{Science} \textbf{\bibinfo{volume}{302}},
  \bibinfo{pages}{89} (\bibinfo{year}{2003}).

\bibitem[{\citenamefont{Rozenberg et~al.}(1999)\citenamefont{Rozenberg, Chitra,
  and Kotliar}}]{rozenberg1999}
\bibinfo{author}{\bibfnamefont{M.~J.} \bibnamefont{Rozenberg}},
  \bibinfo{author}{\bibfnamefont{R.}~\bibnamefont{Chitra}}, \bibnamefont{and}
  \bibinfo{author}{\bibfnamefont{G.}~\bibnamefont{Kotliar}},
  \bibinfo{journal}{Physical review letters} \textbf{\bibinfo{volume}{83}},
  \bibinfo{pages}{3498} (\bibinfo{year}{1999}).

\bibitem[{\citenamefont{Lankhorst et~al.}(2018)\citenamefont{Lankhorst, Poccia,
  Stehno, Galda, Barman, Coneri, Hilgenkamp, Brinkman, Golubov, Tripathi
  et~al.}}]{lankhorst2018}
\bibinfo{author}{\bibfnamefont{M.}~\bibnamefont{Lankhorst}},
  \bibinfo{author}{\bibfnamefont{N.}~\bibnamefont{Poccia}},
  \bibinfo{author}{\bibfnamefont{M.~P.} \bibnamefont{Stehno}},
  \bibinfo{author}{\bibfnamefont{A.}~\bibnamefont{Galda}},
  \bibinfo{author}{\bibfnamefont{H.}~\bibnamefont{Barman}},
  \bibinfo{author}{\bibfnamefont{F.}~\bibnamefont{Coneri}},
  \bibinfo{author}{\bibfnamefont{H.}~\bibnamefont{Hilgenkamp}},
  \bibinfo{author}{\bibfnamefont{A.}~\bibnamefont{Brinkman}},
  \bibinfo{author}{\bibfnamefont{A.~A.} \bibnamefont{Golubov}},
  \bibinfo{author}{\bibfnamefont{V.}~\bibnamefont{Tripathi}},
  \bibnamefont{et~al.}, \bibinfo{journal}{Physical Review B}
  \textbf{\bibinfo{volume}{97}}, \bibinfo{pages}{020504}
  (\bibinfo{year}{2018}).

\bibitem[{\citenamefont{Li et~al.}(2015)\citenamefont{Li, Aron, Kotliar, and
  Han}}]{li2015}
\bibinfo{author}{\bibfnamefont{J.}~\bibnamefont{Li}},
  \bibinfo{author}{\bibfnamefont{C.}~\bibnamefont{Aron}},
  \bibinfo{author}{\bibfnamefont{G.}~\bibnamefont{Kotliar}}, \bibnamefont{and}
  \bibinfo{author}{\bibfnamefont{J.~E.} \bibnamefont{Han}},
  \bibinfo{journal}{Physical Review Letters} \textbf{\bibinfo{volume}{114}},
  \bibinfo{pages}{226403} (\bibinfo{year}{2015}).

\bibitem[{\citenamefont{Stoliar et~al.}(2013)\citenamefont{Stoliar, Cario,
  Janod, Corraze, Guillot-Deudon, Salmon-Bourmand, Guiot, Tranchant, and
  Rozenberg}}]{stoliar2013}
\bibinfo{author}{\bibfnamefont{P.}~\bibnamefont{Stoliar}},
  \bibinfo{author}{\bibfnamefont{L.}~\bibnamefont{Cario}},
  \bibinfo{author}{\bibfnamefont{E.}~\bibnamefont{Janod}},
  \bibinfo{author}{\bibfnamefont{B.}~\bibnamefont{Corraze}},
  \bibinfo{author}{\bibfnamefont{C.}~\bibnamefont{Guillot-Deudon}},
  \bibinfo{author}{\bibfnamefont{S.}~\bibnamefont{Salmon-Bourmand}},
  \bibinfo{author}{\bibfnamefont{V.}~\bibnamefont{Guiot}},
  \bibinfo{author}{\bibfnamefont{J.}~\bibnamefont{Tranchant}},
  \bibnamefont{and}
  \bibinfo{author}{\bibfnamefont{M.}~\bibnamefont{Rozenberg}},
  \bibinfo{journal}{Advanced Materials} \textbf{\bibinfo{volume}{25}},
  \bibinfo{pages}{3222} (\bibinfo{year}{2013}).

\bibitem[{\citenamefont{Teitel and Jayaprakash}(1983)}]{teiteljaya83}
\bibinfo{author}{\bibfnamefont{S.}~\bibnamefont{Teitel}} \bibnamefont{and}
  \bibinfo{author}{\bibfnamefont{C.}~\bibnamefont{Jayaprakash}},
  \bibinfo{journal}{Physical Review Letters} \textbf{\bibinfo{volume}{51}},
  \bibinfo{pages}{1999} (\bibinfo{year}{1983}).

\bibitem[{\citenamefont{Shih and Stroud}(1985)}]{shih1985}
\bibinfo{author}{\bibfnamefont{W.}~\bibnamefont{Shih}} \bibnamefont{and}
  \bibinfo{author}{\bibfnamefont{D.}~\bibnamefont{Stroud}},
  \bibinfo{journal}{Physical Review B} \textbf{\bibinfo{volume}{32}},
  \bibinfo{pages}{158} (\bibinfo{year}{1985}).

\bibitem[{\citenamefont{Newrock et~al.}(2000)\citenamefont{Newrock, Lobb,
  Geigenm{\"u}ller, and Octavio}}]{newrock2000}
\bibinfo{author}{\bibfnamefont{R.}~\bibnamefont{Newrock}},
  \bibinfo{author}{\bibfnamefont{C.}~\bibnamefont{Lobb}},
  \bibinfo{author}{\bibfnamefont{U.}~\bibnamefont{Geigenm{\"u}ller}},
  \bibnamefont{and} \bibinfo{author}{\bibfnamefont{M.}~\bibnamefont{Octavio}},
  \bibinfo{journal}{Solid State Physics} \textbf{\bibinfo{volume}{54}},
  \bibinfo{pages}{263} (\bibinfo{year}{2000}).

\bibitem[{\citenamefont{Granato}(2008)}]{eg08}
\bibinfo{author}{\bibfnamefont{E.}~\bibnamefont{Granato}},
  \bibinfo{journal}{Physical Review Letters} \textbf{\bibinfo{volume}{101}},
  \bibinfo{pages}{027004} (\bibinfo{year}{2008}).

\bibitem[{\citenamefont{Granato}(2013)}]{eg13}
\bibinfo{author}{\bibfnamefont{E.}~\bibnamefont{Granato}},
  \bibinfo{journal}{Phys. Rev. B} \textbf{\bibinfo{volume}{87}},
  \bibinfo{pages}{094517} (\bibinfo{year}{2013}).

\bibitem[{\citenamefont{Marconi and Dom{\'\i}nguez}(2001)}]{marconi2001}
\bibinfo{author}{\bibfnamefont{V.~I.} \bibnamefont{Marconi}} \bibnamefont{and}
  \bibinfo{author}{\bibfnamefont{D.}~\bibnamefont{Dom{\'\i}nguez}},
  \bibinfo{journal}{Physical Review Letters} \textbf{\bibinfo{volume}{87}},
  \bibinfo{pages}{017004} (\bibinfo{year}{2001}).

\bibitem[{\citenamefont{Benz et~al.}(1990)\citenamefont{Benz, Rzchowski,
  Tinkham, and Lobb}}]{benz90}
\bibinfo{author}{\bibfnamefont{S.~P.} \bibnamefont{Benz}},
  \bibinfo{author}{\bibfnamefont{M.~S.} \bibnamefont{Rzchowski}},
  \bibinfo{author}{\bibfnamefont{M.}~\bibnamefont{Tinkham}}, \bibnamefont{and}
  \bibinfo{author}{\bibfnamefont{C.~J.} \bibnamefont{Lobb}},
  \bibinfo{journal}{Phys. Rev. B} \textbf{\bibinfo{volume}{42}},
  \bibinfo{pages}{6165} (\bibinfo{year}{1990}).

\bibitem[{\citenamefont{Jiang et~al.}(2004)\citenamefont{Jiang, Dikin,
  Chandrasekhar, Metlushko, and Moshchalkov}}]{jiang2004}
\bibinfo{author}{\bibfnamefont{Z.}~\bibnamefont{Jiang}},
  \bibinfo{author}{\bibfnamefont{D.}~\bibnamefont{Dikin}},
  \bibinfo{author}{\bibfnamefont{V.}~\bibnamefont{Chandrasekhar}},
  \bibinfo{author}{\bibfnamefont{V.}~\bibnamefont{Metlushko}},
  \bibnamefont{and}
  \bibinfo{author}{\bibfnamefont{V.}~\bibnamefont{Moshchalkov}},
  \bibinfo{journal}{Applied Physics Letters} \textbf{\bibinfo{volume}{84}},
  \bibinfo{pages}{5371} (\bibinfo{year}{2004}).

\bibitem[{\citenamefont{Tripathi et~al.}(2016)\citenamefont{Tripathi, Galda,
  Barman, and Vinokur}}]{tripathi2016}
\bibinfo{author}{\bibfnamefont{V.}~\bibnamefont{Tripathi}},
  \bibinfo{author}{\bibfnamefont{A.}~\bibnamefont{Galda}},
  \bibinfo{author}{\bibfnamefont{H.}~\bibnamefont{Barman}}, \bibnamefont{and}
  \bibinfo{author}{\bibfnamefont{V.~M.} \bibnamefont{Vinokur}},
  \bibinfo{journal}{Physical Review B} \textbf{\bibinfo{volume}{94}},
  \bibinfo{pages}{041104} (\bibinfo{year}{2016}).

\bibitem[{\citenamefont{Jos{\'e} et~al.}(1977)\citenamefont{Jos{\'e}, Kadanoff,
  Kirkpatrick, and Nelson}}]{jose1977}
\bibinfo{author}{\bibfnamefont{J.~V.} \bibnamefont{Jos{\'e}}},
  \bibinfo{author}{\bibfnamefont{L.~P.} \bibnamefont{Kadanoff}},
  \bibinfo{author}{\bibfnamefont{S.}~\bibnamefont{Kirkpatrick}},
  \bibnamefont{and} \bibinfo{author}{\bibfnamefont{D.~R.}
  \bibnamefont{Nelson}}, \bibinfo{journal}{Physical Review B}
  \textbf{\bibinfo{volume}{16}}, \bibinfo{pages}{1217} (\bibinfo{year}{1977}).

\bibitem[{\citenamefont{Franz and Teitel}(1995)}]{franz1995vortex}
\bibinfo{author}{\bibfnamefont{M.}~\bibnamefont{Franz}} \bibnamefont{and}
  \bibinfo{author}{\bibfnamefont{S.}~\bibnamefont{Teitel}},
  \bibinfo{journal}{Physical Review B} \textbf{\bibinfo{volume}{51}},
  \bibinfo{pages}{6551} (\bibinfo{year}{1995}).

\bibitem[{\citenamefont{Lee and Teitel}(1994)}]{leeteitel94}
\bibinfo{author}{\bibfnamefont{J.-R.} \bibnamefont{Lee}} \bibnamefont{and}
  \bibinfo{author}{\bibfnamefont{S.}~\bibnamefont{Teitel}},
  \bibinfo{journal}{Physical Review B} \textbf{\bibinfo{volume}{50}},
  \bibinfo{pages}{3149} (\bibinfo{year}{1994}).

\bibitem[{\citenamefont{Hyman et~al.}(1995)\citenamefont{Hyman, Wallin, Fisher,
  Girvin, and Young}}]{hyman95}
\bibinfo{author}{\bibfnamefont{R.}~\bibnamefont{Hyman}},
  \bibinfo{author}{\bibfnamefont{M.}~\bibnamefont{Wallin}},
  \bibinfo{author}{\bibfnamefont{M.}~\bibnamefont{Fisher}},
  \bibinfo{author}{\bibfnamefont{S.}~\bibnamefont{Girvin}}, \bibnamefont{and}
  \bibinfo{author}{\bibfnamefont{A.}~\bibnamefont{Young}},
  \bibinfo{journal}{Physical Review B} \textbf{\bibinfo{volume}{51}},
  \bibinfo{pages}{15304} (\bibinfo{year}{1995}).

\bibitem[{\citenamefont{Granato and Kosterlitz}(1998)}]{gK98}
\bibinfo{author}{\bibfnamefont{E.}~\bibnamefont{Granato}} \bibnamefont{and}
  \bibinfo{author}{\bibfnamefont{J.}~\bibnamefont{Kosterlitz}},
  \bibinfo{journal}{Physical Review Letters} \textbf{\bibinfo{volume}{81}},
  \bibinfo{pages}{3888} (\bibinfo{year}{1998}).

\bibitem[{\citenamefont{Granato}(1998)}]{eg98}
\bibinfo{author}{\bibfnamefont{E.}~\bibnamefont{Granato}},
  \bibinfo{journal}{Phys. Rev. B} \textbf{\bibinfo{volume}{58}},
  \bibinfo{pages}{11161} (\bibinfo{year}{1998}).

\bibitem[{\citenamefont{Granato}(1996)}]{eg96}
\bibinfo{author}{\bibfnamefont{E.}~\bibnamefont{Granato}},
  \bibinfo{journal}{Physical Review B} \textbf{\bibinfo{volume}{54}},
  \bibinfo{pages}{R9655} (\bibinfo{year}{1996}).

\bibitem[{\citenamefont{Stewart et~al.}(2007)\citenamefont{Stewart, Yin, Xu,
  and Valles}}]{valles07}
\bibinfo{author}{\bibfnamefont{M.~D.} \bibnamefont{Stewart},
  \bibfnamefont{Jr.}}, \bibinfo{author}{\bibfnamefont{A.}~\bibnamefont{Yin}},
  \bibinfo{author}{\bibfnamefont{J.~M.} \bibnamefont{Xu}}, \bibnamefont{and}
  \bibinfo{author}{\bibfnamefont{J.~M.} \bibnamefont{Valles},
  \bibfnamefont{Jr.}}, \bibinfo{journal}{Science}
  \textbf{\bibinfo{volume}{318}}, \bibinfo{pages}{1273} (\bibinfo{year}{2007}).

\bibitem[{\citenamefont{Roy et~al.}(2017)\citenamefont{Roy, Chauhan, Singh,
  Kumar, Jesudasan, Parab, Sensarma, Bose, and Raychaudhuri}}]{roy2017}
\bibinfo{author}{\bibfnamefont{I.}~\bibnamefont{Roy}},
  \bibinfo{author}{\bibfnamefont{P.}~\bibnamefont{Chauhan}},
  \bibinfo{author}{\bibfnamefont{H.}~\bibnamefont{Singh}},
  \bibinfo{author}{\bibfnamefont{S.}~\bibnamefont{Kumar}},
  \bibinfo{author}{\bibfnamefont{J.}~\bibnamefont{Jesudasan}},
  \bibinfo{author}{\bibfnamefont{P.}~\bibnamefont{Parab}},
  \bibinfo{author}{\bibfnamefont{R.}~\bibnamefont{Sensarma}},
  \bibinfo{author}{\bibfnamefont{S.}~\bibnamefont{Bose}}, \bibnamefont{and}
  \bibinfo{author}{\bibfnamefont{P.}~\bibnamefont{Raychaudhuri}},
  \bibinfo{journal}{Physical Review B} \textbf{\bibinfo{volume}{95}},
  \bibinfo{pages}{054513} (\bibinfo{year}{2017}).

\end{thebibliography}

 \newcommand{\noop}[1]{}

\end{document}